\documentclass[aps,twocolumn,reprint,prl]{revtex4-2}
\usepackage{graphicx} % Required for inserting images

\usepackage{hyperref}
\usepackage{xurl}
\hypersetup{breaklinks=true}

\usepackage{amsmath,amssymb}
\usepackage{comment}

\newcommand{\up}{\uparrow}
\newcommand{\dn}{\downarrow}
\newcommand{\av}[1]{\left \langle #1 \right\rangle}

\newcommand{\kv}{\mathbf{k}}

\newcommand{\abs}[1]{\left | #1 \right|}

\renewcommand\Re{\operatorname{Re}}
\renewcommand\Im{\operatorname{Im}}

\usepackage{tikz}

\date{\today}

\usetikzlibrary{shadings}
\usetikzlibrary{decorations.pathreplacing,decorations.markings}
\usetikzlibrary{arrows.meta}
\usetikzlibrary{calc}
\usetikzlibrary{shapes.geometric}

\definecolor{red2}{RGB}{227,26,28}
\definecolor{blue2}{RGB}{69,117,180}
\definecolor{green2}{RGB}{77,175,74}

\begin{document}
\title{Second-order phase transitions and divergent linear response in dynamical mean-field theory}
\author{Erik G. C. P. van Loon }
\affiliation{NanoLund and Division of Mathematical Physics, Physics Department, Lund University, Sweden}

\begin{abstract}
Second-order phase transitions appear as a divergence in one of the linear response functions. For a system of correlated electrons, the relevant divergent response can and does involve many-particle observables, most famously the double occupancy. Generally, evaluating the linear response function of many-particle observables requires a many-particle generalization of the Bethe-Salpeter equation. However, here I show that the divergence of linear response functions in dynamical mean-field theory is governed by a \emph{two}-particle Bethe-Salpeter equation, even for many-particle observables. The reason for this is that the divergence at the second-order phase transition is produced by the self-consistent feedback of the dynamical mean-field.
\end{abstract}

\maketitle

%\section{Introduction}

Electronic correlations lead to a plethora of phases, from metal-insulator transitions~\cite{Imada98} and magnetism~\cite{Hirsch89,Hirsch84} to charge-density waves~\cite{Hirsch84,Hansmann13,currie2023strange}, Wigner crystals~\cite{Wigner34,smolenski2021signatures,li2021imaging}, phase separation~\cite{Kotliar02}, superconductivity~\cite{jiang2019superconductivity,Qin20}, orbital~\cite{Chan09,Pickem21} and bond order~\cite{Dash21,JuliaFarre22}.
Many of these phases are already present in variants of the Hubbard model~\cite{Qin22,Arovas22}. Second-order phase transitions between these correlated phases at finite temperature are of special interest, since divergences occur in the correlation and response functions at these points, associated with the vanishing second derivative of the free-energy functional~\cite{Kotliar00,vanLoon20PRL}. According to the fluctuation-dissipation theorem, the relevant correlation functions are many-particle observables of higher rank than the order parameter itself. For example, for a ferromagnetic or antiferromagnetic transition, the order parameter $\av{n_{i\sigma}}$ is a single-particle operator, while the relevant correlation function $\av{n_{i\sigma}n_{j\sigma}}$ is a two-particle operator.

Dynamical mean-field theory~\cite{Georges96} (DMFT) is a hugely successful approximation for materials with correlated electrons~\cite{kotliar2004strongly}, based on the theoretical~\cite{Metzner89} and experimental~\cite{Tamai19} observation that the electronic self-energy is often predominantly local. This assumption also leads to simplifications at the two-particle level~\cite{Khurana90,Schweitzer91,Jarrell92,Hafermann14}, which have enabled the calculation of dynamical two-particle correlation functions~\cite{Jarrell92,Park11,vanLoon14PRL,Musshoff19,Strand19,Boehnke20,Acharya22,vanLoon2023} according to the Bethe-Salpeter equation. Thus, second-order phase transitions like the metal-insulator transition can be analyzed at the two-particle level using the DMFT Bethe-Salpeter equation
~\cite{BlumerPHD,Kotliar99,Kotliar00,Krien19,Reitner20,vanLoon20PRL,Melnick20,vanLoon22,Kowalski2023thermodynamic}.

However, this analysis of the Bethe-Salpeter equation appears to be limited to the two-particle correlation functions and thus to single-particle order parameters. This excludes the most simple realization of the metal-insulator transition, where the double occupancy $D$ and its response to a change in the Coulomb interaction strength $dD/dU$ are the quantities of interest~\cite{BlumerPHD,Strand11,Kowalski2023thermodynamic}, so the order parameter is a two-particle operator and the divergent correlation-response function involves four-particle operators, whereas the compressibility $d\!\av{n}\!/d\mu$ does not diverge at the critical point of the particle-hole symmetric Hubbard model~\cite{Eckstein07,Springer20,Reitner20}. 

More generally, considering the free energy as a function of $\mu$, $U$ and possible other parameters, thermodynamic stability is a condition on eigenvalues of the second derivative matrix of the free energy~\cite{Kowalski2023thermodynamic}, which can be expressed in terms of mixed response functions like $\partial D/\partial \mu\vert_U$. Finally, in multi-orbital systems, higher-order crystal field and magnetic order parameters~\cite{Santini09,Pourovskii19,Pourovskii20,Pourovskii21PRL} do not always have a representation as a single-particle observable, which follows from the addition rules for angular momentum in many-electron systems.

Thus, it is relevant to study the response of many-particle observables in correlated electron systems, especially with an eye on possible divergences. For the particular case of the double occupancy, Kowalski et al.~\cite{Kowalski2023thermodynamic} have used the Galitskii-Migdal formula to reduce the problem to single-particle objects, but a more general and systematic approach is clearly beneficial. 

Here, I will show that in DMFT the linear response of many-particle correlation functions and especially their divergence is still governed by the usual, \emph{two}-particle Bethe-Salpeter kernel. In fact, the many-particle order parameter and applied field only show up as ``capping stones'' at the end points of the two-particle Bethe-Salpeter equation. Thus, they do not generate the divergence at the second-order transition and their role is restricted to determining if the divergence is picked up in a particular response function. The reason for this remarkable simplification, from many-particle to two-particle physics, can be traced back to the particular form of the DMFT equations, where the self-consistent feedback of the dynamical mean-field is responsible for the second-order phase transition~\cite{BlumerPHD,Kotliar99,Kotliar00,vanLoon20PRL}. On the other hand, going beyond linear response, the two-particle Bethe-Salpeter is no longer sufficient, as expected. 

Consider a general Hubbard model of the form 
\begin{align}
H=\sum_{\text{sites } a,b}\sum_{\alpha\beta} t_{a\alpha,b\beta} c^\dagger_{a\alpha} c_{b\beta}+\sum_{\text{sites }a} H^\text{local}[\{c^\dagger_{a\alpha},c_{a\beta}\}],
\end{align}
where $a$, $b$ are sites in a lattice, $\alpha$ and $\beta$ are orbital labels (which includes spin), $t_{a\alpha,b\beta}$ is the hopping and $H^\text{local}$ is a local Hamiltonian, which is a function of the creation and annihilation operators $c^\dagger_{a\alpha}$, $c_{a\beta}$ on that particular lattice site. The local Hamiltonian includes many-particle terms such as the Coulomb interaction $\frac{1}{2}\sum_{\alpha\beta\gamma\delta} U_{\alpha\beta\gamma\delta} c^\dagger_\alpha c_\beta c^\dagger_\gamma c_\delta$. Here, $t_{a\alpha,b\beta}$ and $H^\text{loc}$ are Hermitian. For a translationally invariant system, $t_{\kv,\alpha\beta}$ denotes the Fourier transform of $t_{a\alpha,b\beta}$ to momentum space. The model is considered at a finite temperature $T=1/\beta$, and factors of $T$ are suppressed in the equations for compactness.

In DMFT, this lattice Hamiltonian is replaced by an auxiliary impurity model with the same local Hamiltonian but with a dynamical hybridization function $\Delta_{\nu,\alpha\beta}$, where $\nu$ is a fermionic Matsubara frequency.  This hybridization might be represented as an (infinite) discrete bath to obtain a Hamiltonian formulation of the impurity, or simply as an action in imaginary time. For now, a hybridization of the form $\Delta_{\tau-\tau',\alpha\beta} c^\dagger_\alpha(\tau) c^{\phantom{\dagger}}_\beta(\tau')$ is used, where $\Delta_\nu$ has been Fourier transformed to imaginary time. The generalization to Nambu space for superconducting phases is discussed at the end. Given a hybridization $\Delta_{\nu,\alpha\beta}$, the auxiliary impurity model can be solved numerically~\cite{Gull11} and its time-ordered expectation values are denoted by $\av{\cdot}$. In particular, DMFT works with the imaginary-time single-particle Green's function $g_{\alpha\beta}(\tau)=\av{c_{\alpha}(\tau) c_\beta^\dagger}$ and its Fourier transform to Matsubara frequency $g_{\nu,\alpha\beta}$. In the following, $t_\kv$, $\Delta_\nu$ and $g_\nu$ are considered as matrices in orbital space, and $\cdot^{-1}$ denotes the matrix inverse in this space. 

The DMFT loop is closed by a prescription to find the hybridization $\Delta_\nu$, the dynamical mean-field, which is given by a set of self-consistency conditions,
\begin{align}
     \forall_\nu: \,\, 0\overset{!}{=} f_\nu(\Delta_\nu, g_\nu) = g_{\nu} - \int d\kv \left[ g^{-1}_{\nu}+\Delta_{\nu} - t_{\kv} \right]^{-1}. \label{eq:sc:f}
\end{align} 
Here, $\int d\kv=1/N_k \sum_\kv$ denotes taking the momentum average, i.e., the local part.
Equation~\eqref{eq:sc:f} is a coupled set of equations because the solution of the auxiliary impurity model $g_{\nu}$ implicitly depends on $\Delta_{\nu'}$ also for $\nu\neq \nu'$.

\emph{Linear response of local observables}
Linear response considers the change of the expectation value of an operator $\hat{B}$ to a small perturbation $H\rightarrow H-A\hat{X}$ of the Hamiltonian, where $A$ is the magnitude of the perturbation and $A$ and $\hat{X}$ are called conjugate variables. Two examples introduced above are the density of orbital $\alpha$, $\hat{n}_\alpha=c^\dagger_\alpha c^{\phantom{\dagger}}_\alpha$ and the double occupancy on orbital $\alpha$, $\hat{D}_\alpha = c^\dagger_{\alpha\up}c^{\phantom{\dagger}}_{\alpha\up}c^\dagger_{\alpha\dn}c^{\phantom{\dagger}}_{\alpha\dn}$, which are conjugate to the chemical potential $\mu$ and Hubbard interaction $U$ acting on that orbital, respectively.
 
As in these examples, and in the spirit of dynamical mean-field theory, I focus here on homogeneous local perturbations, i.e., $\hat{X}=\sum_{\text{sites }i} \hat{X}_i[\{c^\dagger_{i\alpha},c^{\phantom{\dagger}}_{i\beta} \}]$, where $\hat{X}_i$ is a local operator on site $i$ of arbitrary order. Similarly, only site-local observables $\hat{B}$ are considered. In that case, in DMFT, it makes sense to identify~\footnote{In some cases, the inconsistencies inherent in approximate many-body theories~\cite{Vilk97} make this identification of impurity expectation values as the relevant quantity not unique~\cite{vanLoon16}. However, in the example of the double occupancy~\cite{vanLoon16}, using the impurity expectation value guarantees positivity and the Galitskii-Migdal relation, so it is still the most reasonable choice.} the expectation value of the impurity model as the relevant quantity, i.e., $\av{B}=\frac{1}{N_\#} \sum_{\text{sites }i} \av{B_i}=\av{B}^\text{imp}$, which can be measured in the impurity solver. For the linear response to the homogeneous field $A$, the resulting linear change to a local observable $B$ is the same on all sites, i.e., it is a $\mathbf{q}\!=\!\mathbf{0}$ response. More generally~\cite{Fleck98}, it is also possible to consider how $\av{B_b}$ depends on $\hat{A}_a$ for any pair of sites $a$, $b$, and the corresponding $\mathbf{q}$-dependent response function in momentum space. Similarly, since the perturbation is constant in time, the linear response is also assumed to be time-independent and the response function has $\omega=0$. The linear response formalism assumes that no spontaneous symmetry breaking in space or time takes place in response to the field, but second-order phase transitions are visible as a divergent linear response. For single-particle operators $A\hat{X}$ and $\hat{B}$, the DMFT linear response is given by the well-known Bethe-Salpeter equation~\cite{Georges96}. Here, I show that the approach which was previously used to prove the thermodynamic consistency~\cite{vanLoon15} of the DMFT compressibility can also be used to express the linear response of many-particle observerables in simple terms.

\emph{Derivation}
For a local (i.e., impurity) expectation value $\av{B}$, a change in the parameter $A$ of the local Hamiltonian will lead to both direct changes and indirect changes via the DMFT self-consistent field $\Delta$,
\begin{align}
    \frac{d\av{B}}{dA} &=     \frac{\partial \av{B}}{\partial A}\Bigg\vert_{\Delta} + \sum_{\nu'} \frac{\partial \av{B}}{\partial \Delta_{\nu'}}\Bigg\vert_{A} \frac{\partial \Delta_{\nu'}}{\partial A}. 
\end{align}
This requires the calculation of the change of $\Delta$ with respect to $A$, which can be determined from the fact that the DMFT self-consistency equation has to be satisfied both before and after applying the field. Restating the DMFT self-consistency, Eq.~\eqref{eq:sc:f}, in terms of $g^{-1}$ instead of $g$ will lead to more compact equations in the end~\footnote{Using $g^{-1}$ directly has the benefit of giving amputated correlation functions.}.
\begin{multline}
 f\Bigg(g^{-1}\Big[\Delta[A],A\Big],\Delta[A] \Bigg) \\
 = (g^{-1})^{-1} - \int d\kv \left(g^{-1}+\Delta-t_\kv \right)^{-1}.
 \end{multline}
Here, the square brackets denote that the mean-field $\Delta$ depends on $A$ and the inverse of the local Green's function $g^{-1}$ depends on $A$ both directly and via $\Delta[A]$. $f$ is diagonal in $\nu$, so the $\nu$ labels are suppressed to keep the notation compact. 

As stated before, the objects $g^{-1}$, $\Delta$, $t_\kv$ are matrices in orbital space. The derivative of one of these matrices with respect to another matrix is a rank-4 tensor in orbital space. Furthermore, $g$ and $\Delta$ carry a single fermionic frequency, so the derivative $\partial g^{-1}/\partial \Delta$ has two fermionic frequencies, i.e., it is a matrix. It will make sense to interpret these rank-4 orbital tensors as matrices (rank-2 tensors) in a space of orbital pairs, keeping the additional matrix structure in frequency space as well. In this pair space, the usual single-frequency rank-2 orbital objects are vectors. For the matrix inverse in this pair space, the notation $\cdot^{\neg 1}$ is used, while $\cdot^{-1}$ is reserved for the original orbital space. For matrix derivatives, there is the useful identity $\partial(M^{-1})/\partial x = - M^{-1} \left(\partial M/\partial x\right) M^{-1}$.

To satisfy the self-consistency condition after the infinitesimal change in the external field $A$, 
\begin{widetext}
\begin{align}
    \forall_\nu:\,\,\, 0 = \frac{df_\nu}{dA} &= \frac{\partial f_\nu}{\partial \Delta_\nu}\Big|_{g^{-1}} \frac{\partial \Delta_\nu}{\partial A} + \sum_{\nu'} \frac{\partial f_\nu}{\partial g^{-1}_\nu}\Bigg|_{\Delta} \frac{\partial g^{-1}_\nu}{\partial \Delta_{\nu'}} \frac{\partial \Delta_{\nu'}}{\partial A} + \frac{\partial f_\nu}{\partial g^{-1}_\nu}\Bigg|_{\Delta} \frac{\partial g^{-1}_\nu}{\partial A} \Bigg|_{\Delta}, \label{eq:sc:variation} \\
    \forall_\nu:\,\,\, 0 &= \frac{\partial f_\nu}{\partial \Delta_\nu}\Big|_{g^{-1}} \frac{\partial \Delta_\nu}{\partial A} + \sum_{\nu'} \frac{\partial f_\nu}{\partial g^{-1}_\nu} \left(-\delta_{\nu\nu'}\hat{1}-\frac{\partial \Sigma_\nu}{\partial \Delta_{\nu'}} \right) \frac{\partial \Delta_{\nu'}}{\partial A} - \frac{\partial f_\nu}{\partial g^{-1}_\nu}\Bigg|_{\Delta} \frac{\partial \Sigma_\nu}{\partial A}\Bigg|_{\Delta}. \label{eq:scderiv2}
\end{align}
\end{widetext}
Here, $g^{-1}_\nu = i\nu - \Delta_\nu - \Sigma_\nu$ acts as the definition of the impurity self-energy $\Sigma$. The relevant partial derivatives of the self-consistency condition are
 \begin{align}
     \frac{\partial f}{\partial \Delta}\Big|_{g^{-1}} &= \int d\kv \left[g^{-1}+\Delta-t_\kv \right]^{-1} \frac{\partial \Delta}{\partial \Delta}\left[g^{-1}+\Delta-t_\kv \right]^{-1} \notag \\
    &= \int d\kv \, G_\kv G_\kv \equiv \chi^{0,\text{lat}}, \\
    \frac{\partial f}{\partial g^{-1}}\Big|_{\Delta} &= -(g^{-1})^{-1} \frac{\partial g^{-1}}{\partial g^{-1}} (g^{-1})^{-1} + \notag \\
    & \int d\kv \left[g^{-1}+\Delta-t_\kv \right]^{-1} \frac{\partial g^{-1}}{\partial g^{-1}}\left[g^{-1}+\Delta-t_\kv \right]^{-1} \notag \\
    &= \int d\kv \, G_\kv G_\kv  - g g \equiv \chi^{0,\text{lat}} - \chi^{0,\text{imp}}\equiv \tilde{\chi}^0, 
\end{align}
where so-called bubbles of Green's functions are denoted as $\chi^{0}$, these are rank-4 tensors in orbital space. They are diagonal in frequency, since $f$ depends on $g$ and $\Delta$ at the same frequency only. Seen as a bubble, both propagators have the same frequency because the $\omega=0$ response is being considered. In particular, $\chi^{0,\text{lat}}$ is the bubble of lattice Green's functions (at $q=0$, $\omega=0$), $\chi^{0,\text{imp}}$ is the impurity bubble (also at $\omega=0$) and $\tilde{\chi}^0$ is their difference, the non-local part of the bubble. 
The only term connecting different Matsubara frequencies, $\partial \Sigma/\partial \Delta$ is related to the impurity vertex~\cite{vanLoon20PRL} $F$,
\begin{align}
    \frac{\partial \Sigma_\nu}{\partial \Delta_{\nu'}} \equiv  F_{\nu\nu'} g_{\nu'} g_{\nu'}  = F \chi^{0,\text{imp}}.
\end{align}
Note that both $F$ and $\chi^{0,\text{imp}}$ are rank-4 tensors in orbital space, so they are matrices in pair space and their product is the matrix product in pair space, i.e., another rank-4 tensors in orbital space. Diagrammatically, this corresponds to contracting two legs of both objects. 

\begin{figure}
\begin{tikzpicture}

    \draw[left color=red2,right color=blue2, middle color=white] (0,0) -- ++(0.5,0.2) -- ++(0.5,-0.2) -- ++(-0.5,-0.2) -- cycle ;
    \node[circle,fill=red2,inner sep=0,minimum height=0.1cm] at (0,0) {};    
    \node[circle,fill=blue2,inner sep=0,minimum height=0.1cm] at (1,0) {};            

    \node at (1.5,0) {+};
    
    \begin{scope}[shift={(2.,0)}]

    \draw[left color=red2] (0,0) -- ++ (0.5,0.2) -- ++ (0,-0.4) -- cycle;
    \node[circle,fill=red2,inner sep=0,minimum height=0.1cm] at (0,0) {};    
    \draw[-to reversed] (0.5,0.2) -- node[above,font=\footnotesize] {$\nu$}  ++(0.4,0) -- ++ (-0.2,0) ;
    \draw[-to] (0.5,-0.2) -- node[below,font=\footnotesize] {$\nu$}  ++(0.4,0) -- ++ (-0.2,0) ;

    \draw[right color=blue2,left color=white] (1.4,0) -- ++ (-0.5,0.2) -- ++ (0,-0.4) -- cycle;
    \node[circle,fill=blue2,inner sep=0,minimum height=0.1cm] at (1.4,0) {};    

    \end{scope}

    \node at (4,0) {+};
    
    \begin{scope}[shift={(4.5,0)}]

    \draw[left color=red2] (0,0) -- ++ (0.5,0.2) -- ++ (0,-0.4) -- cycle;
    \node[circle,fill=red2,inner sep=0,minimum height=0.1cm] at (0,0) {};    
    \draw[-to reversed] (0.5,0.2) -- node[above,font=\footnotesize] {$\nu'$} ++(0.4,0) -- ++ (-0.2,0) ;
    \draw[-to] (0.5,-0.2) -- node[below,font=\footnotesize] {$\nu'$} ++(0.4,0) -- ++ (-0.2,0) ;

    \draw[fill=black!20] (0.9,-0.2) -- ++(0.4,0) -- ++(0,0.4) -- ++(-0.4,0) -- ++ (0,-0.4) ;
    \draw[-to reversed] (1.3,0.2) -- node[above,font=\footnotesize] {$\nu{\phantom{'}}$}  ++(0.4,0) -- ++ (-0.2,0) ;
    \draw[-to] (1.3,-0.2) -- node[below,font=\footnotesize] {$\nu{\phantom{'}}$}  ++(0.4,0) -- ++ (-0.2,0) ;

    \draw[right color=blue2,left color=white] (2.2,0) -- ++ (-0.5,0.2) -- ++ (0,-0.4) -- cycle;
    \node[circle,fill=blue2,inner sep=0,minimum height=0.1cm] at (2.2,0) {};    

    \end{scope}

    \node at (0.5,-1.5) {+};

    \begin{scope}[shift={(1,-1.5)}]

    \draw[left color=red2] (0,0) -- ++ (0.5,0.2) -- ++ (0,-0.4) -- cycle;
    \node[circle,fill=red2,inner sep=0,minimum height=0.1cm] at (0,0) {};    
    \draw[-to reversed] (0.5,0.2) -- node[above,font=\footnotesize] {$\nu_1$} ++(0.4,0) -- ++ (-0.2,0) ;
    \draw[-to] (0.5,-0.2) -- node[below,font=\footnotesize] {$\nu_1$} ++(0.4,0) -- ++ (-0.2,0) ;

    \draw[fill=black!20] (0.9,-0.2) -- ++(0.4,0) -- ++(0,0.4) -- ++(-0.4,0) -- ++ (0,-0.4) ;
    \draw[-to reversed] (1.3,0.2) -- node[above,font=\footnotesize] {$\nu_2$}  ++(0.4,0) -- ++ (-0.2,0) ;
    \draw[-to] (1.3,-0.2) -- node[below,font=\footnotesize] {$\nu_2$}  ++(0.4,0) -- ++ (-0.2,0) ;
    \draw[fill=black!20] (1.7,-0.2) -- ++(0.4,0) -- ++(0,0.4) -- ++(-0.4,0) -- ++ (0,-0.4) ;
    \draw[-to reversed] (2.1,0.2) -- node[above,font=\footnotesize] {$\nu_3$}  ++(0.4,0) -- ++ (-0.2,0) ;
    \draw[-to] (2.1,-0.2) -- node[below,font=\footnotesize] {$\nu_3$}  ++(0.4,0) -- ++ (-0.2,0) ;
    \draw[right color=blue2,left color=white] (3.0,0) -- ++ (-0.5,0.2) -- ++ (0,-0.4) -- cycle;
    \node[circle,fill=blue2,inner sep=0,minimum height=0.1cm] at (3.0,0) {};    

    \end{scope}    

    \node at (4.5,-1.5) {+};    
    \node at (5,-1.5) {\ldots};

\end{tikzpicture}
\caption{Diagrammatic representation of the linear response. The geometric series is shown up to second order, higher-order terms have additional vertices and particle-hole propagators inserted. Lines with arrows represent the non-local propagator $\tilde{G}$, a pair of these lines represents a nonlocal bubble $\tilde{\chi}^0$.}
\label{fig:diagram}
\end{figure}
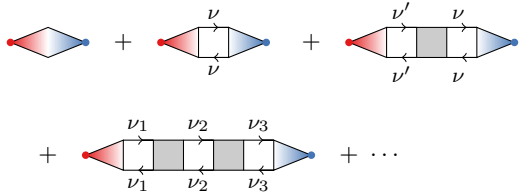

Inserting these results into Eq.~\eqref{eq:scderiv2} and using the pair-frequency space notation (i.e., bubbles and vertices are matrices, derivatives with respect to $A$ are vectors) gives
\begin{align}
    0 &= \sum_{\nu'} \left(\chi^{0,\text{lat}}_{\nu\nu'}  - \tilde{\chi}^0_{\nu\nu'}-\left(\tilde{\chi}^0\frac{\partial \Sigma}{\partial \Delta}\right)_{\nu\nu'} 
    \right)\frac{\partial \Delta_{\nu'}}{\partial A} - \tilde{\chi}^0 \frac{\partial \Sigma_\nu}{\partial A} \notag, \\
    0 &= \sum_{\nu'} \left(\chi^{0,\text{imp}} - \tilde{\chi}^0F \chi^{0,\text{imp}}  
    \right)_{\nu\nu'} \frac{\partial \Delta_{\nu'}}{\partial A} - \tilde{\chi}^0 \frac{\partial \Sigma_\nu}{\partial A} .
\end{align}
Isolating $\partial \Delta/\partial A$ gives 
\begin{align}
&\left(\chi^{0,\text{imp}} - \tilde{\chi}^0 F \chi^{0,\text{imp}} \right) \frac{\partial \Delta}{\partial A} = \tilde{\chi}^0 \frac{\partial \Sigma}{\partial A}, \\
&\frac{\partial \Delta}{\partial A} = \left(\chi^{0,\text{imp}}\right)^{\neg 1}\left(\hat{1} - \tilde{\chi}^0 F \right)^{\neg 1}\tilde{\chi}^0 \frac{\partial \Sigma}{\partial A}, 
\end{align}
with $\hat{1}$ the unit matrix in pair-frequency space. Finally,
\begin{multline}
    \frac{d\av{B}}{dA} =     \frac{\partial \av{B}}{\partial A}\bigg\vert_{\Delta} 
    + \frac{\partial \av{B}}{\partial \Delta}\bigg\vert_{A} \left(\chi^{0,\text{imp}}\right)^{\neg 1}\left(\hat{1} - \tilde{\chi}^0 F \right)^{\neg 1}\tilde{\chi}^0 \frac{\partial \Sigma}{\partial A}    
    \label{eq:linearresponse:final}. 
\end{multline}
Here, $\partial \Sigma/\partial A$ is the connected time-ordered correlator $\av{\mathcal{T} Ac c^\dagger}-\av{A}\av{\mathcal{T}cc^\dagger}$ with the fermionic legs amputated~\cite{vanLoon18}, while $\partial \av{B}/\partial \Delta$ is the connected time-ordered correlator $\av{\mathcal{T} Bcc^\dagger}-\av{B}\av{\mathcal{T}cc^\dagger}$ and $\left(\chi^{0,\text{imp}}\right)^{\neg 1}$ corresponds to amputating both its fermionic legs. Both depend on a single fermionic frequency (and $\omega=0$). Finally, $\partial \av{B}/\partial A$ is the connected time-ordered correlator $\av{\mathcal{T} B A}-\av{B}\av{A}$. The ingredients of equation~\eqref{eq:linearresponse:final} are illustrated in Fig.~\ref{fig:components}, while Fig.~\ref{fig:diagram} contains a diagrammatic representation of Eq.~\eqref{eq:linearresponse:final} itself, where the geometric series $(\hat{1}-\tilde{\chi}^0 F)^{\neg 1}$ has been expanded up to second order in the nonlocal Bethe-Salpeter kernel~\cite{vanLoon20PRL} $\tilde{\chi}^0 F$.

\begin{figure}
\begin{tikzpicture}
    \draw[left color=red2,right color=blue2, middle color=white] (0,0) -- ++(0.5,0.2) -- ++(0.5,-0.2) -- ++(-0.5,-0.2) -- cycle ;
    \node[circle,fill=red2,inner sep=0,minimum height=0.1cm] at (0,0) {};    
    \node[circle,fill=blue2,inner sep=0,minimum height=0.1cm] at (1,0) {};    
    \node[left,red2] at (0,0) {$B$}; 
    \node[right,blue2] at (1,0) {$A$};

    \node[anchor=center] at (0.5,-0.7) {$\partial \av{B}/\partial A$};

%%%%%%%%%%%%%%%%%%%%%%%%%%%%%%%%%%%%%%%
    \begin{scope}[shift={(2.5,0)}]

    \draw[left color=red2] (0,0) -- ++ (0.5,0.2) -- ++ (0,-0.4) -- cycle;
    \node[circle,fill=red2,inner sep=0,minimum height=0.1cm] at (0,0) {};    
    \node[left,red2] at (0,0) {$B$}; 
    \draw[-to reversed] (0.5,0.2) -- ++(0.3,0) -- ++ (-0.15,0) ;
    \draw[-to] (0.5,-0.2) -- ++(0.3,0) -- ++ (-0.15,0) ;

    \node[anchor=center] at (0.25,-0.7) {$\partial \av{B}/\partial \Delta_{\nu}$};

    \end{scope}

%%%%%%%%%%%%%%%%%%%%%%%%%%%%%%%%%%%%%%%
    \begin{scope}[shift={(5,0)}]

    \draw[right color=blue2,left color=white] (0,0) -- ++ (-0.5,0.2) -- ++ (0,-0.4) -- cycle;
    \node[circle,fill=blue2,inner sep=0,minimum height=0.1cm] at (0,0) {};    
    \node[right,blue2] at (0,0) {$A$}; 
    \draw[-to] (-0.5,0.2) -- ++(-0.3,0) -- ++ (0.15,0) ;
    \draw[-to reversed] (-0.5,-0.2) -- ++(-0.3,0) -- ++ (0.15,0) ;

    \node[anchor=center] at (-0.25,-0.7) {$\partial \Sigma_\nu/\partial A$};
    
    \end{scope}

%%%%%%%%%%%%%%%%%%%%%%%%%%%%%%%%%%%%%%%
    \begin{scope}[shift={(6.7,0)}]

    \draw[fill=black!20] (-0.2,-0.2) -- (0.2,-0.2) -- (0.2,0.2) -- (-0.2,0.2) -- cycle ;
    \draw[-to reversed] (0.2,0.2) -- ++(0.3,0) -- ++(-0.15,0) ;
    \draw[-to] (0.2,-0.2) -- ++(0.3,0) -- ++(-0.15,0) ;
    \draw[-to] (-0.2,0.2) -- ++(-0.3,0) -- ++(0.15,0) ;
    \draw[-to reversed] (-0.2,-0.2) -- ++(-0.3,0) -- ++(0.15,0) ;

    \node[anchor=center] at (0,-0.7) {$\partial \Sigma_\nu/\partial \Delta_{\nu'}$};
    
    \end{scope}
    
\end{tikzpicture}
\caption{Diagrammatic representation of contributors to the response. The black lines with arrows indicate fermionic propagators. Note that some of the fermionic propagators are amputated and some are not, see main text for definitions. The operators $A$ and $B$ are denoted by small blue and red dots, respectively.}
\label{fig:components}
\end{figure}
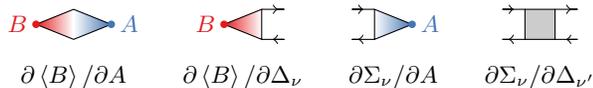

\emph{Second-order phase transitions}
Looking at Eq.~\eqref{eq:linearresponse:final}, none of the impurity correlation functions can be responsible for the divergence, since the impurity model is a finite system at finite temperature, whose expectation values are smooth functions of the model parameters. Instead, the inversion in Eq.~\eqref{eq:linearresponse:final} is the origin of divergences~\cite{vanLoon20PRL}. Exactly at the critical point, one of the eigenvalues of the nonlocal Bethe-Salpeter kernel $\tilde{\chi}^0 F$ is equal to 1, so the inverse in Eq.~\eqref{eq:linearresponse:final} is divergent, and the associated eigenvector $V$ describes the order parameter of the transition. This can be seen as a matrix generalization of the Stoner criterion, where $\tilde{\chi}^0$ describes how many electronic fluctuations are available while $F$ is the effective interaction between correlated electrons.  
 
In the pair-frequency space view, the second term in  Eq.~\eqref{eq:linearresponse:final} is a scalar product of the form vector-matrix-vector. The matrix which is inverted in Eq.~\eqref{eq:linearresponse:final} is independent of $A$ and $B$, i.e., it is always a \emph{two}-particle kernel, even when $A$ and $B$ are many-particle operators, and a single eigenvector $V$ describes the divergent linear response of any observables with respect to any applied field. The overlap between the eigenvector $V$ and the two capping vertices $\partial\av{B}/\partial \Delta$ and $\partial \Sigma/\partial A$ determines if a particular response function $d\!\av{B}\!/dA$ is divergent at the critical point. In particular, symmetry can lead to vanishing overlap, thereby avoiding a divergent response.

The Supplemental Material shows the antiferromagnetic transition~\cite{Sangiovanni06,Schuler18} as an example, where the linear response with respect to the staggered field ($A=h_\text{AF}$) is divergent at $U=U_c$ while the linear response with respect to the interaction strength ($A=U$) is not. The reason is that for the antiferromagnetic transition, the relevant eigenvector $V$ is spin-antisymmetric and thus has non-zero overlap with $\partial \Sigma/\partial h$ which is also spin-antisymmetric, but zero overlap with $\partial \Sigma/\partial U$ which is spin-symmetric.

Another example is the metal-insulator transition in the particle-hole symmetric Hubbard model, where at the critical point  $d\!\av{D}\!/dU$ is divergent~\cite{Strand11,Kowalski2023thermodynamic} but $d\!\av{n}\!/d\mu$ is not~\cite{Eckstein07} because of frequency symmetry~\cite{Springer20,vanLoon20PRL,Reitner20}: the eigenvector $V$ is frequency-antisymmetric while $\partial \Sigma/\partial \mu$ and $\partial \!\av{n}\!/\partial \Delta$ are frequency-symmetric at particle-hole symmetry~\cite{vanLoon14}, so their overlap with $V$ is zero. Below the critical temperature, the resulting hysteresis region has three co-existing solutions (two stable) with different values of $\av{D}$, but a single value of $\av{n}$. 

Physically, the reason for any divergence in DMFT is a runaway self-consistent response of $\Delta$ to an external perturbation, and the self-consistency equation governing $\Delta$ only involves single-particle operators. In linear response, taking first derivatives thus leads to two-particle correlations only, explaining why the two-particle kernel is sufficient. 

On the other hand, the first non-linear response, $d^2\av{B}/dA_1dA_2$, requires a three-particle equivalent of the Bethe-Salpeter equation. It enters through the derivative $\partial^2 \Delta/\partial A_1 \partial A_2$, which can be isolated from a three-particle equivalent of Eq.~\eqref{eq:sc:variation}. This equation will contain a three-particle impurity vertex $\partial^2 g^{-1}/\partial \Delta^2$. Subsequent higher orders require Bethe-Salpeter equations of higher and higher order.

\emph{Superconductivity}
Superconductivity can be described in DMFT and its cluster extensions using the Nambu formulation~\cite{georges1993superconductivity,Lichtenstein00,Harland16,Karp22}, where the dynamical mean-field also has anomalous components of the form $\Delta^\text{an.} (c^\dagger c^\dagger+ c c)$, which leads to anomalous components in $g$ and in all vertices. To find instabilities towards a superconducting phase, it is necessary to take these anomalous processes into account in the Bethe-Salpeter equation, even in the normal phase, where it corresponds to the particle-particle channel of the nonlocal Bethe-Salpeter equation, see Refs.~\cite{Otsuki14,DelRe19} for a recent discussion. For this situation, the present derivation can be generalized by incorporating a Nambu index into the orbital label, which leads to a treatment of the particle-particle and particle-hole channels on equal footing. Diagrammatically, propagators and capping vertices with two incoming or outgoing fermions are then allowed. With this generalization, the conclusions about the nature of second-order transitions in DMFT hold, since the necessary ingredient is that the dynamical mean-field couples to precisely two fermionic operators, regardless of their Nambu index.

%\section{Extended DMFT}
\emph{Extensions}
The so-called extended DMFT~\cite{Si96,Parcollet99,Chitra99,Smith00,Chitra01} (EDMFT) and its generalizations~\cite{Sun02,Rubtsov12,Ayral12,vanLoon14,Ayral15} introduce additional dynamical mean fields which couple to densities or other composite operators instead of individual fermionic operators, e.g., a term $\Lambda(\tau-\tau') n(\tau) n(\tau')$ in the impurity model. This $\Lambda(\omega)$ is determined using a many-particle self-consistency condition similar to Eq.~\eqref{eq:sc:f}, whose variation automatically generates many-particle vertices even when single-particle observables like $dn/d\mu$ are considered~\cite{vanLoon15}. Thus, the two-particle Bethe-Salpeter kernel is generally insufficient to identify second-order transitions in these extensions of DMFT.

\emph{Locality}
The approach presented here is restricted to perturbations and operators which are impurity-local and spatially homogeneous, $\mathbf{q}\!=\!\mathbf{0}$. The generalization to commensurate $\mathbf{q}\!\neq\! \mathbf{0}$ is straightforward~\cite{Fleck98,supp}. On the other hand, an extension to non-local operators, e.g., the linear response to changes in $t_\kv$ or the identification of bond ordering, requires more work. In the same vein, the response to changes in temperature changes the Matsubara frequencies themselves, which requires an extension of the current formalism. 

In conclusion, I have shown that the linear response in dynamical mean-field theory is mainly governed by the two-particle Bethe-Salpeter equation, even when many-particle observables are considered. In fact, the specific form of the applied perturbation and the studied observable only appears as capping vertices at the two ends of the Bethe-Salpeter ladder. This generalizes previous formulas for the (density,double occupancy)-($\mu$,$U$) response matrix~\cite{Kowalski2023thermodynamic} to arbitrary local observables and perturbations. 
The DMFT linear response functions are equivalent to second derivatives of the free energy~\cite{Kotliar00,vanLoon14PRL,Kowalski2023thermodynamic}, so this result shows that any DMFT second-order phase transition or thermodynamic instability must appear in the nonlocal Bethe-Salpeter kernel. Furthermore, the spatial structure of the equation is entirely captured by the nonlocal Bethe-Salpeter kernel, so all divergent response functions have the same correlation length close to the phase transition.

\acknowledgments

The author acknowledges useful discussions with L Pourovskii, M. Reitner, G. Sangiovanni, T. Sch\"afer, H. Strand and A. Toschi. The author acknowledges support from the Crafoord Foundation, the Swedish Research Council (Vetenskapsrådet, VR) under grant 2022-03090 and by eSSENCE, a strategic research area for e-Science, grant number eSSENCE@LU 9:1.
The computations were enabled by resources provided by LUNARC, the Centre for Scientific and Technical Computing at Lund University through the projects LU 2024/2-24 and LU 2023/17-14.

\bibliography{references}

\onecolumngrid
\appendix
\clearpage

\section*{Supplemental Material}

\section*{Case study: antiferromagnetic transition on the cubic lattice}

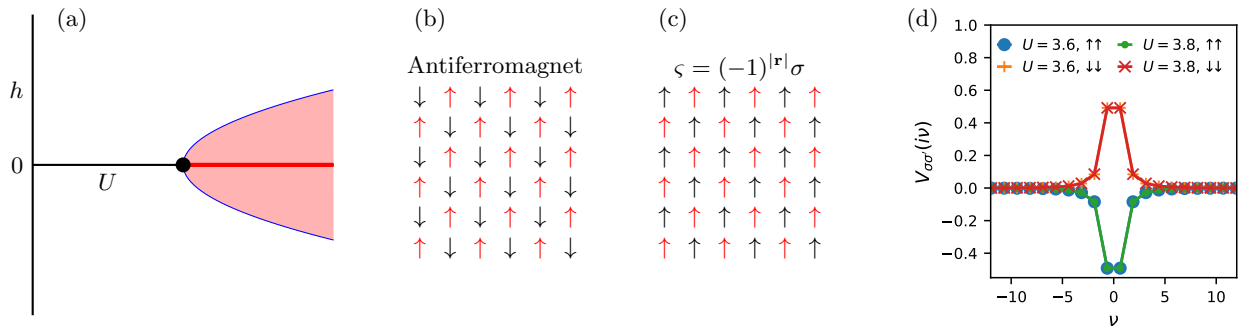
\begin{figure}
\begin{tikzpicture}
 \fill[domain=-1:1,samples=50,fill=red!30] plot ({2+2*\x*\x},\x) --cycle;
 \draw[domain=-1:1,samples=50,blue] plot ({2+2*\x*\x},\x);
 \draw[thick] (0,0) -- node[below,pos=0.25] {$U$} (4,0) ;
 \draw[thick] (0,-2) -- node[pos=0.75,left] {$h$} node[left] {0} (0,2) ;
 \draw[ultra thick,red] (2,0) -- (4,0) ;
 \node[circle,minimum size=2mm,inner sep=0,fill=black] at (2,0) {};
 
 \begin{scope}[shift={(4.75,-1.5)},scale=0.4]
 \node at (3.5, 7) {Antiferromagnet};
\foreach \x in {1,3,5}{
    \foreach \y in {1,3,5}{
        \node[text=red] at (\x,\y) {$\up$} ;
        \node at (\x+1,\y) {$\dn$} ;
        \node at (\x,\y+1) {$\dn$} ;
        \node[text=red] at (\x+1,\y+1) {$\up$} ;
  }
}
 \end{scope}

 \begin{scope}[shift={(8.,-1.5)},scale=0.4]
 \node at (3.5, 7) {$\varsigma=(-1)^{\abs{\mathbf{r}}} \sigma$};
\foreach \x in {1,3,5}{
    \foreach \y in {1,3,5}{
        \node[text=red] at (\x,\y) {$\up$} ;
        \node at (\x+1,\y) {$\up$} ;
        \node at (\x,\y+1) {$\up$} ;
        \node[text=red] at (\x+1,\y+1) {$\up$} ;
  }
}
 \end{scope}

 \node[anchor=north west] at (11.5,2.2) {\includegraphics[width=0.25\textwidth]{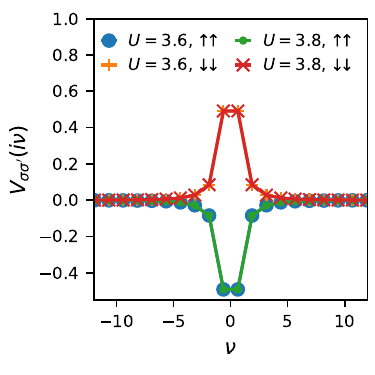}};

 \node[anchor=north west] at (0.2,2.2) {(a)} ;
 \node[anchor=north west] at (4.75+0.2,2.2) {(b)} ;
 \node[anchor=north west] at (8.+0.2,2.2) {(c)} ;
 \node[anchor=north west] at (11.5,2.2) {(d)} ;
 
\end{tikzpicture}%
\caption{The antiferromagnetic transition on a bipartite lattice, in DMFT. (a) Sketch of the phase diagram at fixed temperature, $h$ is the staggered field and $U$ the Hubbard interaction. 
(b-c) Antiferromagnetic order before and after the transformation $\varsigma=(-1)^{\abs{r}} \sigma$. 
(d) Leading eigenvector of the dual Bethe-Salpeter kernel for $U$ just below $U_c$, the corresponding eigenvalue is 0.9 at $U=3.6$ and 0.95 at $U=3.8$.
%This phase diagram is similar to that of the Ising model in mean-field theory.
% In the white region, there is a single solution, while there are three solutions in the red shaded area. Exactly at $h=0$, the red line, two of these solutions with $\abs{m}\neq 0$ are stable while the $\abs{m}=0$ solution is unstable. For $h\neq 0$, there is one stable, one meta-stable and one unstable solution. At the critical point (black dot), the three solutions merge. %At the critical point, $dm/dh$ diverges, i.e., the response in the vertical direction. On the other hand, for a generic observable $X$, $X(U,h)$ is continuous but not smooth at $U_c$, which means that $dX/dU$ is not divergent.
}
\label{fig:sketch}
\end{figure}

At low temperature and large interaction strength, the Hubbard model has a tendency towards strong magnetic fluctuations. On a bipartite lattice at half-filling, such as the cubic lattice, this takes the form of antiferromagnetic order. In both dynamical and static mean-field theory, the transition is characterized by the number of solutions to the self-consistent equations, as shown in Fig.~\ref{fig:sketch}(a). If the staggered field $h=0$ then there is always a solution with staggered magnetization $m=0$. However, additional solutions with finite $\abs{m}$ for $h=0$ appear at $U=U_c$ (black dot) and continue to exist for $U>U_c$ (thick red line). The solution with $m=0$ is stable up to $U_c$ and not beyond. For $h\neq 0$ and $U>U_c$, initially there are multiple DMFT solutions but the one with the staggered magnetization aligned along the field is the unique global minimum of the free energy (red shaded area). For larger $h$, the other solutions eventually disappear (blue line). This phase diagram is similar to mean-field theory for the Heisenberg model. Note that there is a difference with mean-field theory for the Ising model, since there is a continuum of solutions with fixed $\abs{m}$ at $h=0$, instead of a pair $\pm \abs{m}$.

On a bipartite lattice, the staggered field and magnetization can be made uniform by defining a new spin variable $\varsigma=(-1)^\mathbf{|r|} \sigma$, where $\up=-\dn$, as shown in Fig.~\ref{fig:sketch}(b). This mapping leaves the Hubbard interaction invariant, the staggered field changes to $-hc^\dagger_{i,\varsigma} c^{\phantom{\dagger}}_{i,\varsigma}$,  
but the hopping term turns into $-t c^\dagger_{i,\varsigma} c^{\phantom{\dagger}}_{j,-\varsigma}$. This transformation makes it possible to do the DMFT with one single-site impurity model in the ordered phase or in the presence of a staggered field. Even though the hopping is no longer spin-diagonal, the hybridization, local Green's function and self-energy are spin-diagonal even in the presence of the staggered field $h$ since any local process involves an even number of ``hops'' on the bipartite lattice. In the same vein, the system remains at half-filling at $\mu=U/2$ for a bipartite lattice. The mapping to rotated spin variables to make an antiferromagnetic spin pattern uniform can also be generalized to other $\mathbf{q}$~\cite{Fleck98}.

\section*{Numerics without staggered field} 

\begin{figure}
 \includegraphics{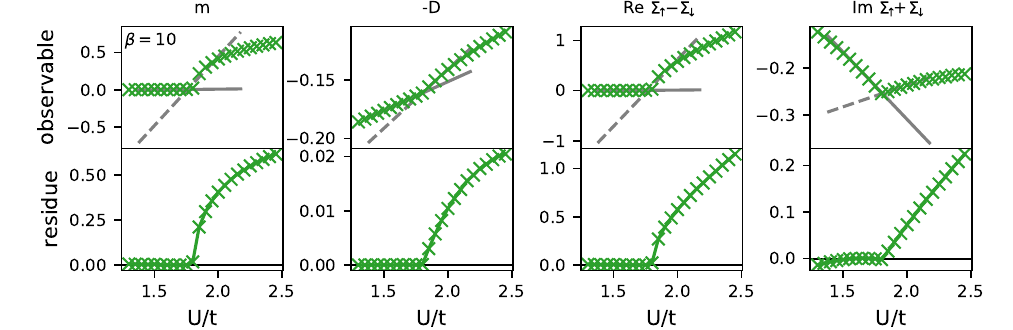}
 \includegraphics{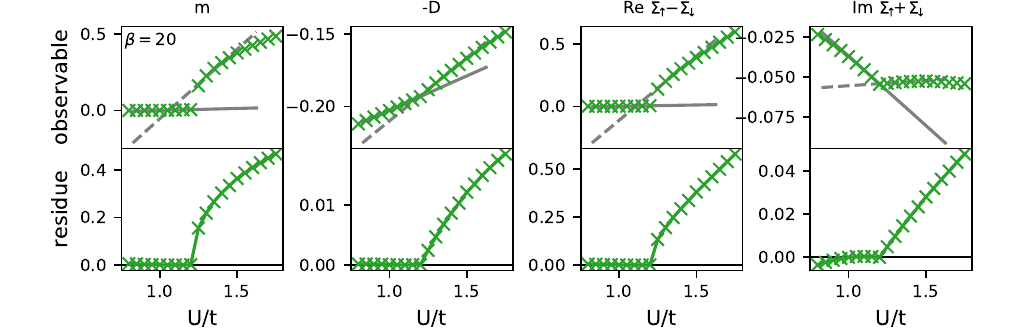}
 \includegraphics{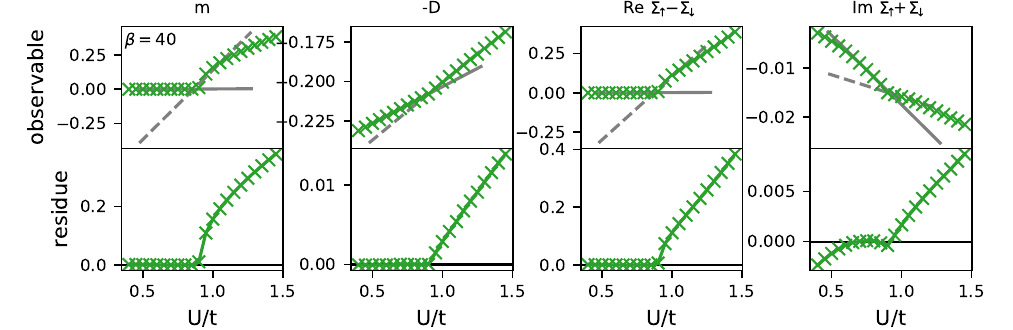}
  \caption{Hypercubic lattice at $h=0$, benchmark data from Ref.~\cite{Schuler18,*zenodo}.}
  \label{suppfig:hypercubic}
\end{figure}

\begin{figure}
 \includegraphics{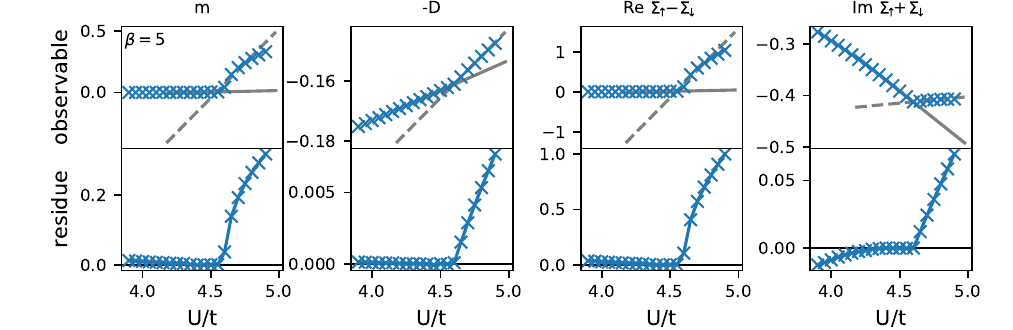}
 \includegraphics{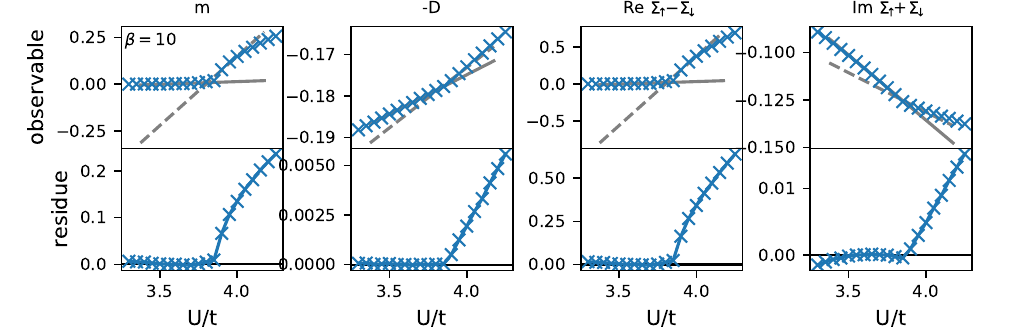}
 \includegraphics{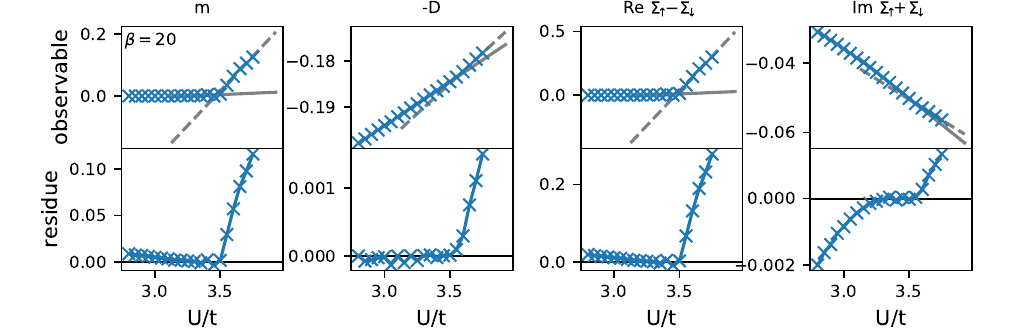}
 \includegraphics{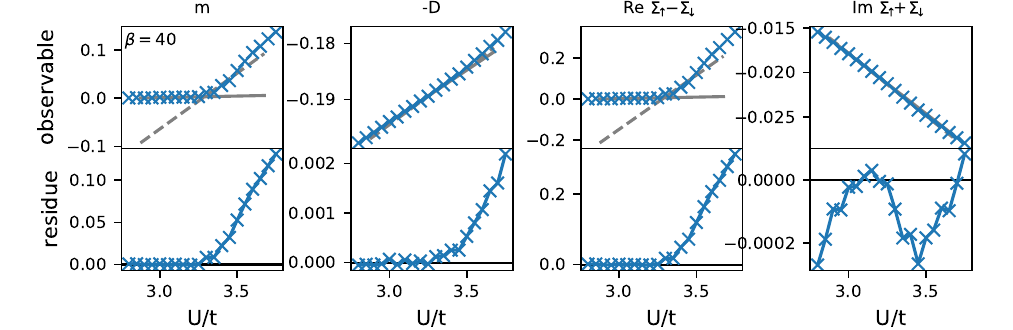}
 \caption{Diamond lattice at $h=0$, benchmark data from Ref.~\cite{Schuler18,*zenodo}.}
\label{suppfig:diamond} 
\end{figure}

\begin{figure}
 \includegraphics{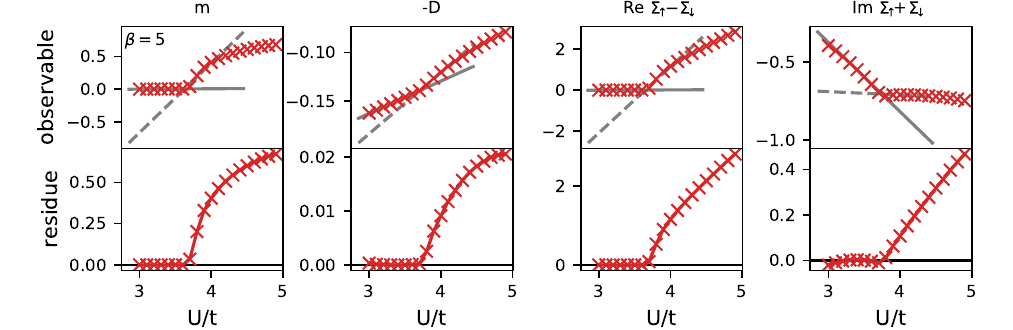}
 \includegraphics{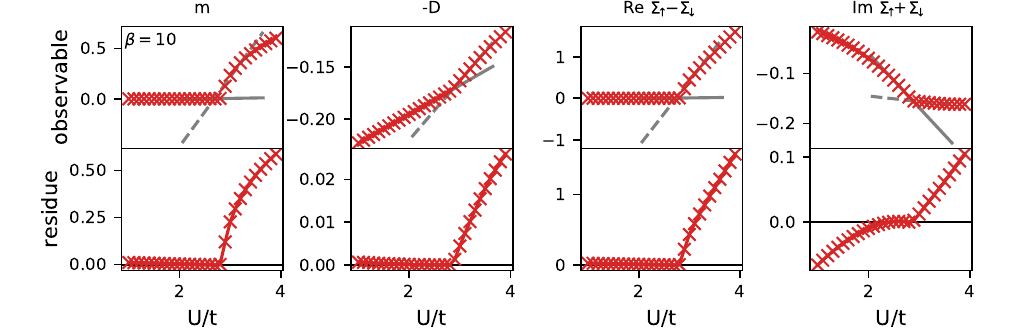}
 \includegraphics{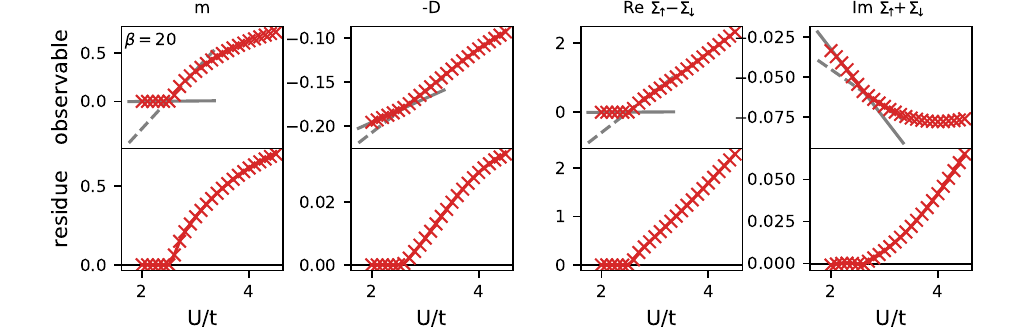}
 \includegraphics{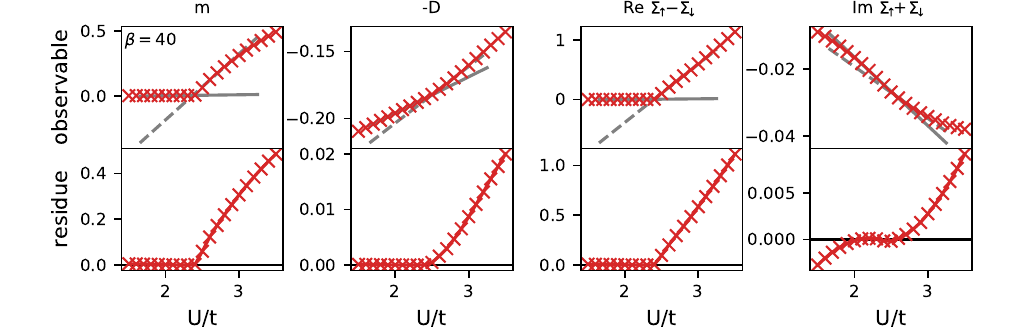}
 \caption{Hyperdiamond lattice at $h=0$, benchmark data from Ref.~\cite{Schuler18,*zenodo}.}
\label{suppfig:hyperdiamond}
\end{figure}

The transition has been studied numerically in previous works~\cite{Sangiovanni06} and there is a publicly available dataset~\cite{Schuler18,*zenodo} for the case $h=0$ for the cubic lattice~\footnote{For completeness, results for the hypercubic, diamond and hyperdiamond lattice are shown in Figs.~\ref{suppfig:hypercubic},~\ref{suppfig:diamond} and ~\ref{suppfig:hyperdiamond}, respectively.}, shown in Fig.~\ref{suppfig:cubic}. The $U_c$ at which a finite staggered magnetization $m$ appears decreases with the inverse temperature $\beta$, we concentrate on the case $\beta=5$, $U_c \approx 4$. The leading eigenvalue of the dual Bethe-Salpeter kernel approaches unity when $U$ approaches $U_c$ from below, signalling the increasing strength of the self-consistent feedback in DMFT. Figure~\ref{fig:sketch}(d) shows the corresponding eigenvector $V_{\sigma\sigma'}(\nu_n)$ for $U=3.6$ (eigenvalue 0.9) and $U=3.8$ (eigenvalue 0.95). The eigenvector lives in the same vector space as single-particle Green's functions, but does not have all of the properties of a Green's function (no causality guarantee, arbitrary normalization and complex phase), since it denotes possible changes in the Green's function. In the present system, the phase can be chosen so that the eigenvector is entirely real. From Fig.~\ref{fig:sketch}(d), it is clear that the leading eigenvector for the antiferromagnetic transition is spin-antisymmetric, i.e., $V_{\sigma\sigma'}(\nu)=\sigma \delta_{\sigma\sigma'}V(\nu)$ and frequency-symmetric, $V(\nu)=V(-\nu)$. The eigenvector is non-zero for all Matsubara frequencies. The eigenvector apparently does not change much between $U=3.6$ and $U=3.8$, even though the eigenvalue gets substantially closer to unity. In addition to the eigenvector with $S^z$ symmetry shown here, by $SU(2)$ symmetry there is a second eigenvector of the Bethe-Salpeter with the same eigenvalue, which corresponds to spontaneous order with in-plane magnetization. This eigenvector is of the form $V_{\sigma\sigma'}=V_{\up\dn}-V_{\dn\up}$ and is therefore also spin-antisymmetric. Since our finite field calculations have $h$ along the $z$-axis, we focus on the $V_{\sigma\sigma'}=V_{\up\up}-V_{\dn\dn}$ eigenvector here.

\section*{Interaction strength response}
\begin{figure}
 \includegraphics{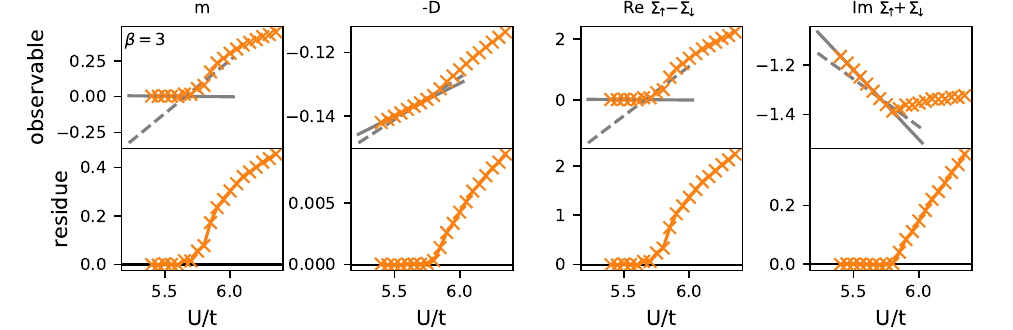}
 \includegraphics{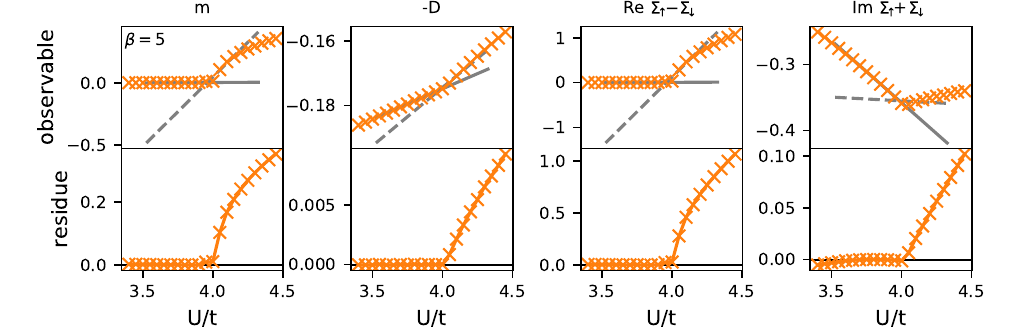}
 \includegraphics{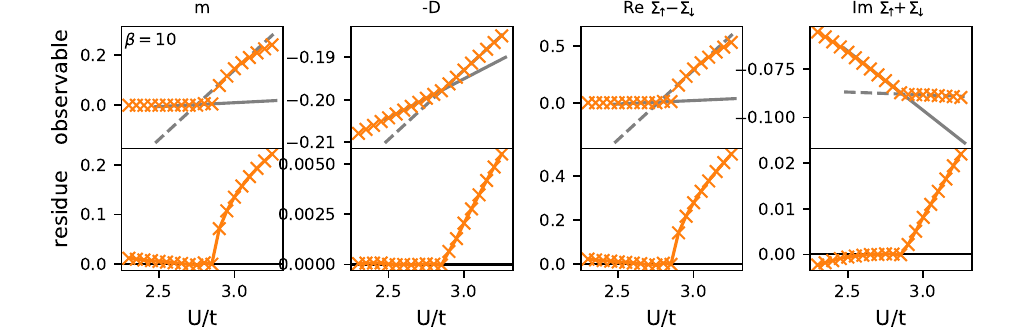}
 \includegraphics{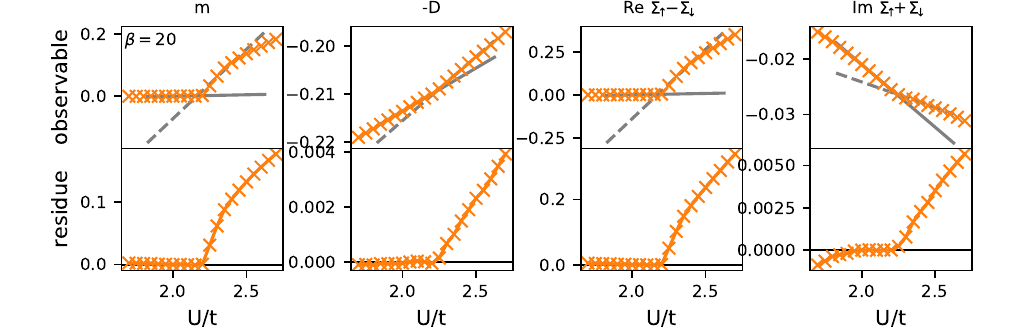}
 \caption{Cubic lattice at $h=0$, benchmark data from Ref.~\cite{Schuler18,*zenodo}.}
 \label{suppfig:cubic}
\end{figure}

We can now consider the response along two directions in parameter space, namely changes in $U$ while keeping $h=0$ or an applied field $h$ at some constant $U$, either at $U<U_c$ or $U>U_c$. We start with the latter case. Figure~\ref{suppfig:cubic} shows additional (dynamical) local observables in addition to the magnetization, all at $h=0$. Shown are the double occupancy $D=n_{i,\up} n_{i,\dn}$, the difference in the real part of the self-energy between majority and minority, $\Re(\Sigma_\up-\Sigma_\dn)$, evaluated at the lowest Matsubara frequency $\nu_0=\pi/\beta$ and the average of the imaginary part of the majority and minority self-energy, $\Im(\Sigma_\up+\Sigma_\dn)$, also evaluated at the lowest Matsubara frequency. We observe that all four observables are continuous but not smooth at $U_c$. The solid lines show fits to the final data points in the disordered phase ($m=0$), while the dashed lines are fits to the first data points in the disordered phase ($\abs{m}>0$). In the bottom panels of Fig.~\ref{suppfig:cubic}, the disordered phase fit (solid line) is subtracted, to more clearly show the non-smooth behavior around $U_c$. Since all shown observables $B$ are continuous functions of $U$, the linear response with respect to the interaction, $dB/dU$, is not divergent, but the non-linear response $d^2 B/dU^2$ can be divergent. 

This can be understood based on the analysis in the main text, since the capping vertex $\partial \Sigma/\partial U|_\Delta$ and its overlap with the leading eigenvector of the non-local Bethe-Salpeter equation decides if the linear response is divergent~\footnote{The linear response formula also contains the bubbles $\chi^{0,\text{imp}}$ and $\tilde{\chi}^0$, which are frequency-dependent renormalization factors that do no change the analysis of the spin symmetry.}. Here, approach the critical point from the disordered phase with $h=0$ and $U<U_c$, $\partial \Sigma/\partial U|_\Delta$ is spin-symmetric, so the overlap with the spin-antisymmetric leading eigenvector is zero. Thus, there is no divergent linear response of any observable with respect to $U$.

%\begin{figure}
% \includegraphics{fig_hscan_U3.500.pdf}
% \includegraphics{fig_hscan_U5.000.pdf}
% \caption{Self-energy as a function of antiferromagnetic field $h$. Data for $h<0$ obtained by symmetry from $\Sigma_{\downarrow}$ at $h>0$. Cubic lattice, half-filling, $\beta=5$ and $t=1$.} 
% \label{suppfig:hscan}
%\end{figure}

\begin{figure} 
 \includegraphics{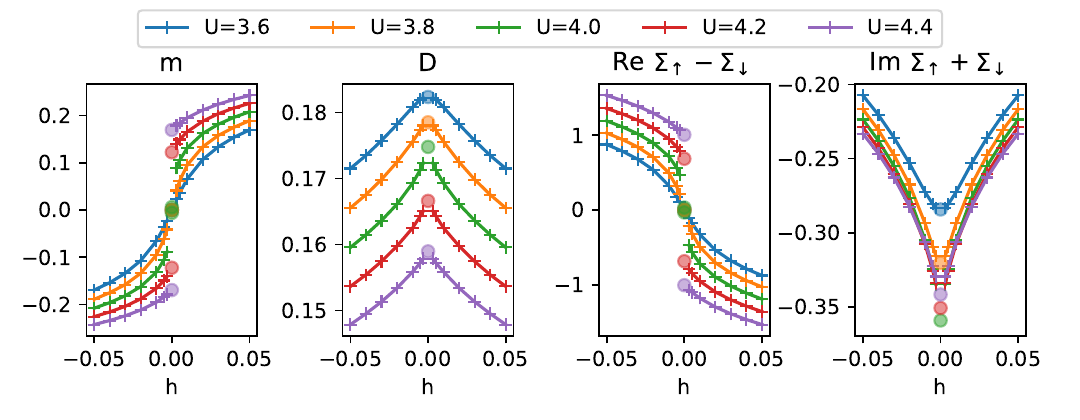} 
 \caption{The impact of an applied staggered field $h$. DMFT for the Hubbard model on the cubic lattice at $\beta=5$ ($U_c\approx 4$). Data for $h<0$ obtained from $h>0$ by symmetry, the circles indicate the $h=0$ benchmark~\cite{Schuler18,*zenodo}, see also Fig.~\ref{suppfig:cubic}. For $\Sigma$, the lowest Matsubara frequency $\nu_0=\pi T$ is shown. Note that at every $h$, only solutions that are a global minimum of the free are shown. With the additional metastable and unstable solutions, continuous $S$-shaped curves or loops would form, see Ref.~\cite{Strand11}.}  
 \label{suppfig:hscan2}
\end{figure}
%For $U<U_c$, the magnetization $m$ and $\Re \Sigma(i\nu_0)$ show linear response with respect to $h$, while the double occupancy $D$ and $\Im \Sigma(i\nu_0)$ are quadratic in $h$. At $U>U_c$, the functions are no longer smooth, the linear response is replaced by a discontinuity while the quadratic behavior gets a derivative discontinuity. 

\begin{figure} 
 \includegraphics{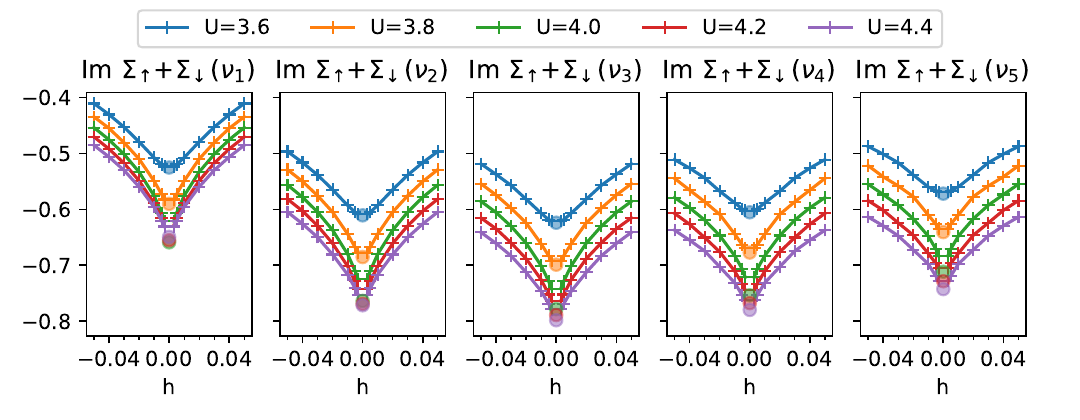}  
 \caption{Frequency-dependence of the self-energy in the presence of an applied staggered field $h$, $\nu_n=(2n+1)\pi T$. DMFT for the Hubbard model on the cubic lattice at $\beta=5$ ($U_c\approx 4$). Data for $h<0$ obtained from $h>0$ by symmetry, the circles indicate the $h=0$ benchmark~\cite{Schuler18,*zenodo}. }  
 \label{suppfig:hscan3} 
\end{figure}

\section*{Field response}
To see the diverging linear response, we have to look at the effect of an applied staggered field $h$, i.e., the vertical direction in Fig.~\ref{fig:sketch}(a). In that case, the capping vertex for the applied field, $\partial \Sigma/\partial h$, has a spin-antisymmetric and frequency-symmetric component already at the Hartree level, so there is a non-zero overlap with the leading eigenvector. Figure~\ref{suppfig:hscan2} shows finite-field results. Close to the phase transition at $U_c\approx 4$, the linear response of all shown observables has a divergence, but there are two qualitatively different cases due to the symmetry of the observables. Starting with the $h$-antisymmetric observables $m$ and $\Re(\Sigma_\up-\Sigma_\dn)$
%$\sum_{\varsigma=\up,\dn} \varsigma \Re \Sigma_\varsigma$,
we find the expected linear response at $U<U_c$, which turns into a discontinuous jump for $U>U_c$. This jump should be understood in the spirit of a Maxwell construction, where an $S$-shaped curve would appear if additional solutions were also shown. At higher Matsubara frequencies (not shown), $\Re \Sigma$ behaves qualitatively similar, although there are quantitative difference because the DMFT self-energy is a dynamic object. At large frequency, $\Re \Sigma$ and $\av{m}$ are related by the asymptotic expression $\lim_{\nu\rightarrow\infty} \Re \Sigma_\varsigma(\nu) = U n_{-\varsigma}$.
%The $h$-symmetric quantities $D$ and $\sum_{\varsigma=\up,\dn} \Im \Sigma_\varsigma$ and the $h$-antisymmetric $m$ and $\sum_{\varsigma=\up,\dn} \varsigma \Re \Sigma_\varsigma$ show clearly different behavior at the phase transition.

For the $h$-symmetric observables $D$ and $\Im \Sigma(\nu_0)$, the field-dependence is quadratic and smooth for $U<U_c$. For $U>U_c$, there is a derivative discontinuity at $h=0$. If the metastable and unstable solutions would also be drawn, the curves would form a loop, as in Ref.~\cite{Strand11}. As for the real part, other frequencies of $\Im \Sigma(\nu_n)$ are qualitatively similar, as is visible in Fig.~\ref{suppfig:hscan3}. Going to high frequency, the asymptotic relation $\lim_{\nu\rightarrow\infty} \Im \Sigma_\varsigma =\frac{U^2}{i\nu} \av{n_{-\varsigma}}(1-\av{n_{-\varsigma}})$ at half-filling simplifies using $\av{n_{-\varsigma}}(1-\av{n_{-\varsigma}})=(1-m^2)/4$, so $\Im \Sigma$ is continuous but not smooth at $h=0$ because the same holds for $m^2$. 

The divergent response of these observables is explained by the capping vertex $\partial \av{B}/\partial \Delta|_{h=0}$ and its spin symmetry. That is, for all four shown observables, their impurity value changes if $\Delta$ changes along the direction given by the eigenvector in Fig~\ref{fig:sketch}(d). Here, it is worth pointing out that we are looking at $B=\Re \Sigma(\nu)$ and $B=\Im \Sigma(\nu)$ one Matsubara frequency at a time, it is possible to get zero response instead of a divergence if one looks at a linear combination like $\Re( \Sigma(\nu)-\Sigma(-\nu))$, because of the usual symmetry of $\Sigma$. In terms of the Bethe-Salpeter analysis, this is because the corresponding capping vertex of this linear combination is zero by symmetry.

\section*{Comparison to Mott transition}

For the non-magnetic Mott metal-insulator transition in DMFT~\cite{Georges96,Strand11,vanLoon20PRL,Reitner20}, the leading eigenvector is spin-symmetric, i.e., $V_{\sigma\sigma'}(\nu)=\delta_{\sigma\sigma'} V(\nu)$ and frequency-antisymmetric, $V(-\nu)=-V(\nu)$. In that case, the response with respect to changes in the interaction strengh is divergent at the critical point, but the response with respect to the chemical potential is not divergent~\cite{Eckstein07}. For the metal-insulator transtion, the lack of divergence in $d\av{n}/d\mu$ is explained by the frequency symmetry of the eigenvector~\cite{Springer20}, instead of the spin symmetry. 

An important difference is that the order parameter of the antiferromagnetic transition corresponds to a clear spontaneous symmetry breaking. As a result, the line $h=0$ in the phase diagram is special and everything is (anti)symmetric around this line. The first-order transition takes place exactly at $h=0$ for $U>U_c$. For the Mott transition in the $U$-$T$ plane, there is less symmetry, so the first-order transition line $U_c(T)$ in the hysteresis region is a curve instead of a straight line. 

In principle, for the antiferromagnetic transition, there could be observables $B$ for which $d\!\av{B}\!/dh$ is neither zero nor divergent at the critical point, similar to the finite $d(\frac{dn}{d\mu})/dU$ for the metal-insulator transition. This would happen if the capping vertex $d\av{B}/d\Delta$ has non-zero overlap with $d\Sigma/dh$, zero overlap with the leading eigenvector of the non-local Bethe-Salpeter equation, but non-zero overlap with other eigenvectors with eigenvalue smaller than 1. The first condition requires a spin-antisymmetric component, so then the second requirement means that the overlap with $V$ should vanish due to the frequency structure (instead of the spin structure). This did not happen for any observables studied here, but in principle one can construct by hand linear combination of observables where this happens. In multi-orbital Hubbard models, the larger number of reasonable observables makes this scenario more likely.

\section*{Simulation details}

The finite-field DMFT simulations shown here were performed using TRIQS~\cite{triqs}, the TRIQS CTHYB solver~\cite{cthyb}. Close to a second order phase transition, the convergence of the DMFT self-consistency cycle slows down and many iterations are needed~\cite{vanLoon20PRL}, here approximately 100 iterations were used. The unit of energy $t=1$ is used. Finite field calculations were performed for $B=0.003,0.005,0.01,0.02,0.03,0.04$ and $0.05$. 
The eigenvectors in Fig.~\ref{fig:sketch}(d) are calculated without applied field, using w2dynamics~\cite{w2dynamics} for the vertex that enters the Bethe-Salpeter equation and TPRF~\cite{Strand:tprf} for the evaluation of the Bethe-Salpeter equation.

\end{document}